\documentclass[11pt]{article}
\usepackage{CJK}
\usepackage{graphicx}
\usepackage{fancyhdr}
\usepackage{amsmath,amsthm,amssymb}
\usepackage{mathrsfs}
\usepackage{multirow}
\usepackage{indentfirst}
\usepackage{color}
\usepackage{epstopdf}
\usepackage{url}

\newtheorem{theorem}{Theorem}
\newtheorem{remark}{Remark}
\newtheorem{lemma}{Lemma}

\numberwithin{equation}{section}
\makeatletter
\newcommand{\Rmnum}[1]{\expandafter\@slowromancap\romannumeral #1@}
\makeatother

\allowdisplaybreaks

\begin{document}
\date{}

\title{\Large \bf Testing heteroscedasticity for regression models based on projections
\footnote{Corresponding author:   jiangxj@sustc.edu.cn (Xuejun Jiang).}}
{\small \author{Falong Tan$^{1}$, Xuejun Jiang$^{2*}$, Xu Guo$^3$ and Lixing Zhu$^{3, 4}$ \\
{\small {\small {\it $^1$ College of Finance and Statistics, Hunan University, China} }}\\
{\small {\small {\it $^2$ Department of Mathematics, South University of Science and Technology of China, China } }}\\
{\small {\small {\it $^3$ School of Statistics, Beijing Normal University, China} }}\\
{\small {\small {\it $^4$ Department of Mathematics, Hong Kong Baptist University, Hong Kong} }}\\
}}
\date{}
\maketitle

{\small
\noindent {\bf Abstract:} In this paper we propose a new test of heteroscedasticity for parametric regression models and partial linear regression models in high dimensional settings. When the dimension of covariates is large, existing tests of heteroscedasticity perform badly due to the ``curse of dimensionality". To attack this problem, we construct a test of heteroscedasticity by using a projection-based empirical process.
We study the asymptotic properties of the test statistic under the null hypothesis and alternative hypotheses. It is shown that the test can detect local alternatives departure from the null hypothesis at the fastest possible rate in hypothesis testing.
As the limiting null distribution of the test statistic is not distribution free, we propose a residual-based bootstrap. The validity of the bootstrap approximations is investigated. We present some simulation results to show the finite sample performances of the test. Two real data analyses are conducted for illustration.

\bigskip

}

\newpage

\section{Introduction}
In many regression models the error terms are assumed to have common variance. Ignoring the presence of heteroscedasticity in regression models may result in inefficient inferences of the regression coefficients, or even inconsistent estimators of the variance function. Therefore, testing heteroscedasticity in regression models should be conducted when the error terms are assumed to have equal variance. Consider the following regression model:
\begin{equation}\label{1.1}
Y=m(Z)+\varepsilon,
\end{equation}
where $Y$ is the dependent variable with a $p$-dimensional covariate $Z$, $m(\cdot)=E(Y|Z=\cdot)$ is the regression function, and the error term $\varepsilon$ satisfies $E(\varepsilon|Z)=0$. Thus the null hypothesis in testing heteroscedasticity for the regression model~(\ref{1.1}) is that
\begin{eqnarray*}
H_0: Var(Y|Z)=E(\varepsilon^2|Z) \equiv C \quad {\rm for \ some \ constant} \ C >0;
\end{eqnarray*}
while the alternative hypothesis is that $H_0$ is totally incorrect:
$$ H_1:  Var(Y|Z)=E(\varepsilon^2|Z) \ {\rm is \ a \ nonconstant \ function \ of } \ Z.$$

Testing heteroscedasticity for the regression model~(\ref{1.1}) has been studied by many authors in the literature. Cook and Weisberg (1983) constructed a score test for heteroscedasticity in parametric regression models with parametric structure variance functions.  Dette and Munk (1998) proposed a consistent test for heteroscedasticity in a nonparametric regression setting based on a $L^2$-distance between the underlying variance function and the constant variance. Zhu, Fujikoshi and Naito (2001) developed a test of heteroscedasticity based on empirical processes. Built on the work of Zheng (1996) for checking the regression function, Dette (2002) and Zheng (2009) respectively proposed two residual based tests for heteroscedasticity under different regression models. Lin and Qu (2012) developed a test of heteroscedasticity for nonlinear semi-parametric regression models based on the work of Dette (2002). Furthermore, Dette, Neumeyer and van Keilegom (2007) consider a more general problem of checking the parametric form of the conditional variance function in nonparametric regressions. For newly developed procedures for heteroscedasticity in nonparametric regression models, see for instance, Chown and M\"{u}ller (2018) and Pardo-Fern\'{a}ndez and Jim\'{e}nez-Gamero (2018).

To motivate the construction of our test statistic in this paper, we first give a detailed comment on two representative tests: Zhu, Fujikoshi and Naito (2001)'s test and Zheng (2009)'s test. Let $E(\varepsilon^2)=\sigma^2$ and $\eta=\varepsilon^2-\sigma^2$. Then the null hypothesis $H_0$ is tantamount to $E(\eta|Z)=0$. Consequently,
$$E[\eta E(\eta|Z) f(Z)]=0,$$
where $f(\cdot)$ is the density function of $Z$. Based on a consistent estimator of $E[\eta E(\eta|Z) f(Z)]$, Zheng (2009) proposed a test statistic as follows:
$$ T_n=\frac{1}{n(n-1)} \sum_{i=1}^n \sum_{j \neq i}^n \frac{1}{h^p} K(\frac{X_i-X_j}{h}) \hat{\eta}_i \hat{\eta}_j, $$
where $\hat{\eta}_i=\hat{\varepsilon}_i^2-\hat{\sigma}^2, \hat{\sigma}^2= (1/n) \sum_{i=1}^n \hat\varepsilon^2_i$, $ \hat{\varepsilon}_i=Y_i -\hat{m}(Z_i)$ with $\hat{m}(\cdot)$ being an estimator of the regression function, $K(\cdot)$ is a $p$-dimensional multivariate kernel function and $h$ is a bandwidth. Note that Zheng (2009) used nonparametric smooth estimators to construct the test statistic. Thus Zheng (2009)'s test suffers severely from the ``curse of dimensionality". More specifically, Zheng (2009) can only detect the local alternatives that converge to the null at a rate of $O(1/\sqrt{nh^{p/2}})$, where $p$ is the dimension of $Z$. When $p$ is large, this rate could be very slow and the power of Zheng (2009)'s test drop quickly.

Zhu, Fujikoshi and Naito (2001) use residual marked empirical process to construct a test of heteroscedasticity. Note that
$$ E(\eta|Z)=0 \Leftrightarrow E[\eta I(Z \leq t)]=0 \ {\rm for \ all } \ t \in \mathbb{R}^p.  $$
Based on this, Zhu, Fujikoshi and Naito (2001) proposed a residual marked empirical process as follows:
$$ R_n(t)= \frac{1}{\sqrt{n}} \sum_{i=1}^n \hat{\eta}_i I(Z_i \leq t). $$
The test statistic of Zhu, Fujikoshi and Naito (2001) is a functional of $R_n(t)$ such as the Cram\'{e}r-von Mises or Kolmogorov-Smirnov functional. It is shown that the test of Zhu, Fujikoshi and Naito (2001) can detect the local alternatives converging to the null at the parametric rate $1/\sqrt{n}$ which is the fastest convergence rate in hypothesis testing. But when the dimension $p$ of the covariates is large, this test also suffers severely from the dimension problem due to the data sparseness in high dimensional spaces.

Therefore, how to attack the ``curse of dimensionality" is very important for heteroscedasticity testing. The purpose of this paper is to develop a test of heteroscedasticity in parametric regression models that is suitable for the case in which the dimension of covariates is relatively high. To this end, we use projected covariates $\alpha^{\top}Z$ to construct a residual marked empirical process and the test statistic is a functional of the projected empirical process.
Escanciano (2006) and Lavergne and Patilea (2008, 2012) also adapted this approach to construct goodness-of-fit tests for parametric regression models. As the test is based on one-dimensional projections, it behaves as if the dimension of covariates was one. Thus this method is less sensitive to the dimension $p$ of the regressors than that in Zhu, Fujikoshi and Naito (2001) and Zheng (2009). As the proposed test is based on projected empirical processes, it is able to detect local alternatives converging to the null at the parametric rate. Besides, the new test is easy to compute, does not involve high dimensional numerical integrations, and presents an excellent power performance for large dimension in finite sample simulations, see Section 4.

We also use this method to check heteroscedasticity in partial linear regression models. This is an important issue in high dimensional data analysis. When the dimension of covariate is large, nonparametric estimation is less accurate due to the ``curse of dimensionality'', and partial linear regression models provide a more flexible substitution if the researchers already know some of the covariates enter the regression model linearly. Thus this model is widely used in economics, biology and other related fields.
To construct the test statistic, we need to use locally smoothing methods to estimate the nonlinear part of the regression function in a partial linear regression model. Although it involve nonparametric estimators, we will show that the limiting distribution is the same as that in parametric regression models and the proposed test can also detect local alternatives converging to the null at a rate $1/\sqrt{n}$ under this semi-parametric setting.


The rest of the paper is organized as follows. In section 2 we define the test statistic by using a projection-based empirical process. In section 3 we study the asymptotic properties of the test statistic under the null and alternative hypotheses in parametric regression models and partial linear regression models, respectively. In section 4, a residual-based bootstrap method is proposed to approximate the null distribution of the test statistic, simulation results comparing the proposed test with some existing competitor in the literature are presented, and two real data analyses are used as an illustration. Section 5 contains a discuss. Appendix contains the regularity conditions and technical proofs.


\section{Test construction}
Recall that the null hypothesis $H_0$ is equivalent to $E(\eta|Z)=0$. According to Lemma 1 of Escanciano (2006) or Lemma 2.1 of Lavergne and Patilea (2008), we have
$$ E(\eta|Z)=0 \Longleftrightarrow E(\eta|\alpha^{\top} Z)=0, \quad \forall \ \alpha \in \mathbb{S}^p, $$
where $\mathbb{S}^p=\{\alpha: \alpha \in \mathbb{R}^p \ {\rm and } \ \|\alpha\|=1 \}$. Consequently,
$$E(\eta|Z)=0 \Longleftrightarrow E[\eta I(\alpha^{\top} Z \leq t)]=0, \quad \forall \ \alpha \in \mathbb{S}^p, \ t \in \mathbb{R}. $$
Therefore, the null hypothesis $H_0$ is tantamount to
\begin{equation}\label{2.1}
\int_{\mathbb{S}^p} \int_{\mathbb{R}} |E[\eta I(\alpha^{\top} Z \leq t)]|^2 F_{\alpha}(dt) d\alpha=0,
\end{equation}
where $F_{\alpha}$ is the cumulative distribution function of $\alpha^{\top}Z$ and $d\alpha$ is the uniform density on $\mathbb{S}^p$. Then we propose a test statistic for checking heteroscedasticity of model~(\ref{1.1}) as
\begin{equation}\label{2.2}
HCM_n= \int_{\mathbb{S}^p} \int_{\mathbb{R}} \frac{1}{n} |\sum_{j=1}^{n} \hat{\eta}_j I(\alpha^{\top}Z_j \leq t)|^2 F_{n,\alpha}(dt) d\alpha,
\end{equation}
where $F_{n, \alpha}$ is the empirical distribution function of the projected covariates $\{ \alpha^{\top}Z_j, 1\leq j \leq n \}$.

Note that the test statistic $HCM_n$ involves a high-dimensional integral for large $p$. Indeed, by some elementary calculations,
\begin{eqnarray*}
HCM_n &=& \frac{1}{n} \sum_{i,j=1}^{n} \hat{\eta}_i \hat{\eta}_j \int_{\mathbb{S}^p} \int_{\mathbb{R}} I(\alpha^{\top}Z_i \leq t)I(\alpha^{\top}Z_j \leq t) F_{n,\alpha}(dt)d\alpha  \\
&=& \frac{1}{n^2} \sum_{i,j,k=1}^{n} \hat{\eta}_i \hat{\eta}_j \int_{\mathbb{S}^p} I(\alpha^{\top}Z_i \leq \alpha^{\top}Z_k)I(\alpha^{\top}Z_j \leq \alpha^{\top}Z_k) d\alpha
\end{eqnarray*}
It is well known that high-dimensional integrations are extremely difficult to handle. However, the following Lemma enable us to avoid the high-dimensional integrations and obtain an analytic expression of the test statistic $HCM_n$. It can be found in Appendix B of Escanciano (2006).
\begin{lemma}\label{lemma 1}
Let $u_1, u_2 \in \mathbb{R}^p$ be two non-zero vectors and $\mathbb{S}^p$ be the $p$-dimensional unit sphere. Then we have
$$ \int_{\mathbb{S}^p} I(\alpha^{\top} u_1 \leq 0) I(\alpha^{\top} u_2 \leq 0) d\alpha = \frac{\pi -<u_1, u_2>}{2\pi}, $$
where $d\alpha$ is the uniform density on $\mathbb{S}^p$ and $<u_1, u_2>$ is the angle between $u_1$ and $u_2$.
\end{lemma}

\begin{remark}
The above procedure works for any regression model. In this paper we only deal with parametric regression models and partial linear regression models. It can also be applied to nonparametric regression models. Then we need to use nonparametric methods to estimate the unknown regression function. When the dimension of covariates increases, the behavior of nonparametric estimators quickly deteriorates and thus the resulted test may provide inaccurate statistical inferences in practice. Therefore, how to deal with the dimension problem in testing heteroscedasticity for nonparametric regression models is still a challenging problem.
\end{remark}

\section{Asymptotic results}
First we consider a parametric regression model:
\begin{equation}\label{3.1}
Y=m(Z, \beta) + \varepsilon, \quad  E(\varepsilon|Z)=0,
\end{equation}
where $\beta \in \mathbb{R}^d$ and the regression function $m(\cdot, \beta)$ is known and twice differential with respect to $\beta$. Let $\hat \beta$ be the nonlinear least squares estimator of $\beta$ and $\hat \varepsilon_i=Y_i-m(Z_i, \hat \beta)$. Then
$ \hat{\eta}_i=\hat{\varepsilon}_i^2-\hat{\sigma}^2=\hat{\varepsilon}_i^2 -(1/n)\sum_{i=1}^n \hat\varepsilon^2_i $.
Define the empirical process
$$ V_n(\alpha, t)=\frac{1}{\sqrt n} \sum_{i=1}^{n} \hat{\eta}_i I(\alpha^{\top}Z_i \leq t).$$
Then the test statistic becomes
$$HCM_n= \int_{\mathbb{S}^p} \int_{\mathbb{R}} |V_n(\alpha, t)|^2 F_{n,\alpha}(dt) d\alpha. $$
The following theorem presents the asymptotic properties of $V_n(\alpha, t)$ and then of the test statistic $HCM_n$ in~(\ref{2.2}) under the null hypothesis.

\begin{theorem}\label{theorem 1}
Assume the regularity conditions A1-A2 hold. Under $H_0$, we have
$$ V_n(\alpha, t) \longrightarrow V_{\infty}(\alpha,t) \quad  {\rm in \ distribution},$$
where $V_{\infty}(\alpha, t)$ is a zero-mean Gaussian process with a covariance function
$$K\{(\alpha_1, t_1), (\alpha_2, t_2)\}= E\{ \eta^2 [I(\alpha_1^{\top}Z \leq t_1) - F_{\alpha_1}(t_1)][I(\alpha_2^{\top}Z \leq t_2) - F_{\alpha_2}(t_2)] \}.$$
Furthermore,
$$ HCM_n \longrightarrow  \int_{\mathbb{S}^{p+1}} \int_{\mathbb{R}} V_{\infty}(\alpha, t)^2 F_{\alpha}(dt) d\alpha \quad {\rm in \ distribution}. $$
\end{theorem}

Next we apply the above approach to check heteroscedasticity in partial linear regression models. Consider
\begin{equation}\label{3.2}
Y=\beta^{\top}X + g(T) +\varepsilon, \quad E(\varepsilon|X, T)=0
\end{equation}
where $T\in \mathbb{R}$, $\beta \in \mathbb{R}^q$ is an unknown parameter vector, and $g(\cdot)$ is an unknown smooth function. For this semi-parametric setting, we will show that the conclusions in Theorem 1 continue to hold. However, its proof becomes much more complicated. Since the function $g(\cdot)$ is unknown, it has to be estimated in a nonparametric way. Thus, in the theoretical investigations, the decomposition of the proposed empirical process would involve an U-process. With the help of the theory of U-process in the literature, see, e.g. Nolan and Pollard (1987), we can obtain the same asymptotical property as in Theorem 1 for partial linear regression models.

We now use the kernel method to give the estimators of $\beta$ and $g(\cdot)$. Note that
$$ Y-E(Y|T) =\beta^{\top}[X-E(X|T)] +\varepsilon. $$
Let $\tilde{Y}=Y-E(Y|T)$ and $\tilde{X}=X-E(X|T)$. Then it is easy to see that
$$ \beta=[E \tilde{X} \tilde{X}^{\top}]^{-1} E(\tilde{X} \tilde{Y}). $$
Let $\{(X_i, T_i, Y_i) \}_{i=1}^n$ be an i.i.d. sample from the distribution of $(X,T,Y)$. The resulting estimator of $\beta$ is given by
\begin{equation}\label{3.3}
\hat{\beta}=\left(\frac{1}{n} \sum_{i=1}^{n} [X_i-\hat{E}(X|T_i)][X_i-\hat{E}(X|T_i)]^{\top} \right)^{-1} \left( \frac{1}{n} \sum_{i=1}^{n} [X_i-\hat{E}(X|T_i)][Y_i-\hat{E}(Y|T_i)] \right),
\end{equation}
where
\begin{eqnarray*}
\hat{E}(X|T_i)&=&\frac{1}{n} \sum_{j=1,j \neq i}^{n} X_j K_h(T_i-T_j)/\hat{f}_i(T_i),\\
\hat{E}(Y|T_i)&=&\frac{1}{n} \sum_{j=1,j \neq i}^{n} Y_j K_h(T_i-T_j)/\hat{f}_i(T_i),
\end{eqnarray*}
and $\hat{f}_i(T_i)=(1/n)\sum_{j=1,j \neq i}^n K_h(T_i-T_j)$. Here $K_h(t)=(1/h)K(t/h)$ and $K(\cdot)$ is a kernel function satisfying the conditions in Appendix. Next we consider the estimator of $g(\cdot)$. Note that $g(T)=E(Y-\beta^{\top}X|T)$. Then we can obtain the estimator of $g(T)$ as
\begin{equation}\label{3.4}
\hat{g}(T_i)=\frac{1}{n} \sum_{j=1,j \neq i}^n [Y_j-\hat{\beta}^{\top}X_j]K_h(T_i-T_j)/\hat{f}_i(T_i).
\end{equation}
Under the regularity conditions in Appendix, we can derive the following result.

\begin{lemma}\label{lemma 2}
Under the regularity conditions B1-B4 in Appendix, we have
\begin{equation}\label{3.5}
\sqrt{n}(\hat{\beta}-\beta)= [E \tilde{X} \tilde{X}^{\top}]^{-1} \frac{1}{\sqrt{n}} \sum_{i=1}^{n} \tilde{X} \varepsilon_i + O_p(\frac{1}{\sqrt{n}h}+\sqrt{n}h^2)^{1/2}.
\end{equation}
\end{lemma}
Lemma 2 can be found in Zhu and Ng (2003). Then we can obtain the asymptotic properties of $HCM_n$ in parametric regression models. Let $p=q+1$, $Z=(X^{\top}, T)^{\top}$, and the residual $\hat \varepsilon_i=Y_i-\hat{\beta}^{\top}X_i-\hat{g}(T_i)$. Then the proposed empirical process and the test statistic have same form as before,
\begin{eqnarray*}
V_n(\alpha, t)&=&\frac{1}{\sqrt n} \sum_{i=1}^{n} \hat{\eta}_i I(\alpha^{\top}Z_i \leq t), \\
HCM_n &=& \int_{\mathbb{S}^p} \int_{\mathbb{R}} |V_n(\alpha, t)|^2 F_{n,\alpha}(dt) d\alpha.
\end{eqnarray*}

\begin{theorem}\label{theorem 2}
Under the null hypothesis $H_0$ and the regularity conditions B1-B4 in Appendix, the results in Theorems 1 continue to hold.
\end{theorem}

It is worth mentioning that we make no assumptions about the variance function in partial linear models. This is different from existing tests of heteroscedasticity for partial linear models in the literature. Existing tests usually assumed that the variance function $Var(Y|X, T)$ only depend on $T$.  Under this assumption, we can construct a much simpler test using the covariates $T$, rather than the projected covariates $\alpha^{\top}(X^{\top}, T)^{\top}$. If $Var(Y|X, T)$ is a function of $T$, it follow that $Var(Y|X, T)=E(\varepsilon^2|T)$. Then the null hypothesis $H_0$ is tantamount to $E(\eta|T)=0$. The resulting test statistic is given as follows,
$$ CM_n^1= \int_{\mathbb{R}} |\frac{1}{\sqrt n} \sum_{i=1}^{n} \hat{\eta}_i I(T_i \leq t)|^2 dt. $$
More generally, if $T \in \mathbb{R}^d$ is a multiple random variable, we also encounter the dimension problem for large $d$. Thus we can use the projected covariates $\alpha^{\top}T$ to construct a test of heteroscedasticity. The test statistic becomes
$$  CM_n^2= \int_{\mathbb{S}^d} \int_{\mathbb{R}} |\frac{1}{\sqrt n} \sum_{i=1}^{n} \hat{\eta}_i I(\alpha^{\top} T_i \leq t)|^2F_{n,\alpha}(dt) d\alpha, $$
where $F_{n,\alpha}$ is the empirical distribution function of projected covariates $\{\alpha^{\top}T_i: i=1, \cdots, n\}$. The limiting distributions of $CM_n^1$ and $CM_n^2$ are similar as that of $HCM_n$ we derive here.

Now we investigate the sensitivity of the proposed test to alternative hypotheses. Consider a sequence of local alternatives converging to the null at a convergence rate $c_n$
\begin{equation}\label{3.6}
H_{1n}: E(\varepsilon^2|Z)= \sigma^2 + c_n s(Z),
\end{equation}
where $s(Z)$ is not a constant function of $Z$ with $E[s(Z)]=0$ and $E[s(Z)^2] < \infty$. The following Theorem shows that the test is consistent against all global alternative hypotheses and it can detect the local alternatives converging to the null at a rate up to $1/\sqrt{n}$.

\begin{theorem}\label{theorem 3}
Suppose that the regularity conditions in Theorem 1 or Theorem 2 hold. Then \\
(1) under the alternatives $H_{1n}$ with $\sqrt{n} c_n \to \infty$, we have $ HCM_n \rightarrow \infty $ in probability; \\
(2) under the alternatives $H_{1n}$ with $c_n=1/\sqrt{n}$, we have
$$ HCM_n \longrightarrow  \int_{\mathbb{S}^p} \int_{\mathbb{R}} [V_{\infty}(\alpha, t)+S(\alpha, t)]^2 F_{\alpha}(dt) d\alpha \quad {\rm in \ distribution}, $$
where $S(\alpha, t)=E\{s(Z)[I(\alpha^{\top}Z \leq t)- F_{\alpha}(t)]\}$ is a non-random shift term.
\end{theorem}

The proofs of Theorem 1-3 are presented in Appendix. These theorems confirm the claims we made in the Introduction.  Note that our test can be viewed as a generalization of Zhu, Fujikoshi and Naito (2001)'s test. When the dimension of covariate is one, the proposed test reduces to Zhu, Fujikoshi and Naito (2001)'s test. Thus they share some common desirable feathers: both of them are consistent for all global alternatives; the convergence rate does not relate to the dimension of covariates; they can detect local alternatives of order $1/\sqrt{n}$, regardless of the type of the regression function. Furthermore, we use the projection of covariates rather than covariates to construct the test statistic in this paper. As the test is based on one-dimensional projections, it can alleviate the impact of the dimensionality problem largely. The simulation results in the next section validate these results.





\section{Numerical studies}
\subsection{Simulation studies}
In this subsection we conduct several simulation studies to investigate the performance of the proposed tests. Since the tests are not distribution-free, we suggest a residual-based bootstrap to approximate the distributions of the test statistics. This method has been previously adopted by Hsiao and Li (2001), Wang and Zhou (2007),  Su and Ullah (2013), Guo et al. (2018). The procedure of the residual-based bootstrap is given as follows:
\begin{itemize}
\item[(1).] For a given random sample $\{(Y_i, Z_i): i=1, \cdots, n\}$, obtain the residual $\hat{\varepsilon}_i=Y_i -\hat{m}(Z_i)$ with $\hat{m}(\cdot)$ being an estimator of the regression function.
\item[(2).] Obtain the bootstrap error $\varepsilon_i^{\ast}$ by randomly sampling with replacement from the center variables $\{ \hat{\varepsilon}_i-\bar{\hat \varepsilon}: i=1, \cdots, n \}$ where $\bar{\hat \varepsilon}=(1/n)\sum_{i=1}^n \hat{\varepsilon}_i$. Then define $Y_i^{\ast}=\hat{m}(Z_i)+\varepsilon_i^{\ast}$.
\item[(3).] For the bootstrap sample $\{(Y_i^{\ast}, Z_i): i=1, \cdots, n \}$, obtain the estimator $\hat{m}^{\ast}(Z_i)$ and then define the bootstrap residual $\hat{\varepsilon}^{\ast}_i=Y^{\ast}_i-\hat{m}^{\ast}(Z_i)$. Let $\hat{\eta}^{\ast}_i=\hat{\varepsilon}^{\ast 2}_i-\hat{\sigma}^{\ast 2}_i$ and $\hat{\sigma}^{\ast 2}_i=(1/n)\sum_{i=1}^n \hat{\varepsilon}^{\ast 2}_i$. Thus the bootstrap test statistic $ HCM_n^{\ast}$ is calculated based on $\{(\hat{\eta}^{\ast}_i, Z_i): i=1, \cdots, n \}$.
\item[(4).] Repeat step (2) and (3) a large number of times, say, $B$ times. For a given significant level $\gamma$, the critical value is determined by the upper $\gamma$ quantile of the bootstrap distribution $\{HCM_{n,j}^{\ast}: j=1, \cdots, B\}$ of the test statistic.
\end{itemize}
Note that $\hat{m}(Z_i)=m(Z_i, \hat{\beta})$ for a parametric regression model~(\ref{3.1}) and $\hat{m}(Z_i)=\hat{\beta}^{\top}X_i + \hat{g}(T_i)$ with $Z_i=(X_i, T_i)$ for a partial linear regression model~(\ref{3.2}). The bootstrap estimator $\hat{m}^{\ast}(Z_i)$ is defined similarly.

To establish the validity of the proceeding residual-based bootstrap, we need the following theorem.
\begin{theorem}\label{Theorem 4}
Suppose the regularity conditions in Theorem 1 or Theorem 2 hold. Then\\
(1) under the null $H_0$ and the local alternative $H_{1n}$, the distribution of $HCM_n^{\ast}$ given $\{(Y_i, Z_i): i=1, \cdots, n\}$ converges to the limiting null distribution of $HCM_n$ in Theorem 1. \\
(2) under the alternative $H_1$, the distribution of $HCM_n^{\ast}$ given $\{(Y_i, Z_i): i=1, \cdots, n\}$ converges to a finite limiting distribution.
\end{theorem}

Theorem 4 indicates that the proceeding bootstrap is asymptotically valid. Under the null hypothesis, the bootstrap distribution gives an asymptotically approximation to the limiting null distribution of $HCM_n$. Under the local alternatives $H_{1n}$ and the global alternative $H_1$, the test based on the bootstrap critical values are still consistent.

Next we report some simulation results to show the finite sample performances of the proposed tests. We also make a comparison with Zhu, Fujikoshi and Naito (2001)'s test $T_n^{ZFN}$, Zheng (2009)'s test $T_n^{ZH}$ and Guo et al. (2018)'s test $T_n^{G}$ under different setting of dimensions. Note that  Guo et al. (2018) used characteristic functions to construct a test of heteroscedasticity, which is also based on one-dimensional projections. Thus their test is also less sensitive to the dimension of covariates. The test statistic of Guo et al. (2018) is given as follows,
$$T_n^{G}=\frac{1}{n(n-1)} \sum_{i=1}^n \sum_{j\neq i}^n \hat{\eta}_i \hat{\eta}_j \exp(-\| X_i-X_j \|^{\delta}). $$

In the following examples, $a=0$ corresponds to the null hypothesis and $a \neq 0$ to the alternative hypotheses. The sample sizes are 100 and 200. The empirical sizes and powers are calculated through 1000 replications at a nominal level 0.05. The number of the bootstrap sample is set to be $B=500$. We choose $\delta=1.5$ in $T_n^{G}$, as suggested in Guo et al. (2018).

{\bf Example 1}. The data are generated from the following parametric regression models:
\begin{eqnarray*}
H_{11}: Y&=&\beta^{\top}Z + |a \times \beta^{\top}Z +0.5| \times \varepsilon; \\
H_{12}: Y&=&\beta^{\top}Z + \exp(a \times \beta^{\top}Z) \times \varepsilon; \\
H_{13}: Y&=&\beta^{\top}Z + |a \times \sin(\beta^{\top}Z) +1| \times \varepsilon; \\
H_{14}: Y&=&\exp(-\beta^{\top}Z)+|a \times \beta^{\top}Z +0.5| \times \varepsilon;
\end{eqnarray*}
where $Z \sim N(0, I_p)$, independent of the standard normal error $\varepsilon$ and $\beta=(1, \cdots, 1)^{\top}/\sqrt{p}$. To show the impact of the dimension, $p$ is set to be 2, 4, and 8 in each model. Note that model $H_{13}$ is a high frequency model and the other three are low frequency models. To see whether the regression function can infect the performance of the tests, we consider a nonlinear regression function in model $H_{14}$.

The simulation results are reported in Tables 1-2. It can be observed that when $p=2$, Zheng (2009)'s test $T_n^{ZH}$ and Guo et al. (2018)'s test $T_n^{G}$ can not maintain the significance level for some cases, while the other two tests perform better. For the empirical power, all these tests have high power. But the proposed test $HCM_n$ and Zhu, Fujikoshi and Naito (2001)'s test $T_n^{ZFN}$ grow faster than the other two as $a$ increases. When the dimension $p$ becomes large, the tests $HCM_n$ and $T_n^{ZFN}$ can still control the empirical size for large $p$.  In contrast, the empirical sizes of $T_n^{ZH}$ and $T_n^{G}$ are slightly away for the significant level. For the empirical power, the tests $HCM_n$ and $T_n^{G}$ works much better than the other two and $T_n^{ZFN}$ becomes the worst one as it almost has no empirical powers when $p=8$. These phenomena validate the theoretical results that the proposed test $HCM_n$ is little affected by the dimension of covariates and the tests $T_n^{ZH}$ and $T_n^{ZFN}$ suffer severely from the dimensionality. In the high frequency model $H_{13}$, we can observe that the locally smoothing test $T_n^{ZH}$ performs much worse than the other tests. This is different from the case in model checking where locally smoothing tests usually perform better than globally smoothing tests in high frequency models. Further, we found no significant difference in empirical sizes and powers from different regression functions in models $H_{11}$ and $H_{14}$.
$$\rm  Tables \ 1-2 \ are \ about \ here $$

In the next simulation examples we further investigate the performance of the proposed test in partial linear regression models. We focus on two different cases: (1) $Var(\varepsilon|X, T)$ is a function of $(X,T)$ and (2) $Var(\varepsilon|X, T)$ is a function of $T$.

{\bf Example 2}. The data are generated from the following models:
\begin{eqnarray*}
H_{21}: Y&=&\beta^{\top}X + T^2 + |a (\beta^{\top}X+T) +0.5| \times \varepsilon; \\
H_{22}: Y&=&\beta^{\top}X + T^2 + \exp\{a (\beta^{\top}X+T)\} \times \varepsilon; \\
H_{23}: Y&=&\beta^{\top}X + T^2 + |a \sin(\beta^{\top}X+ T) +1| \times \varepsilon;\\
H_{24}: Y&=&\beta^{\top}X + T^2 + \exp(4 a T) \times \varepsilon;
\end{eqnarray*}
where $X \sim N(0, I_q)$, $T \sim U(0, 1)$, $\varepsilon \sim N(0,1)$ and $\beta=(1, \cdots, 1)^{\top}/\sqrt{q}$. The error term $\varepsilon$ is independent of $(X, T)$. The dimension $q$ of covariates $X$ is also set to be 2, 4 and 8.

We use the kernel function $K(u)=(1/\sqrt{2\pi})\exp(- u^2/2)$. To investigate the impact from the choice of the bandwidth $h$, we consider several values of $h$ in a considerable wide range and then empirically choose one as the bandwidth. Let $h=j/100$ for $j=10, 15, 20, \cdots, 100$. The empirical sizes and powers for different dimensions are presented in Figure 1 and 2.
$$\rm  Figures \ 1-2 \ is \ about \ here $$

From Figure 1 and 2, we can see that when the bandwidth $h$ is too small, $HCM_n$ can not maintain the significant level. When the bandwidth $h$ is large than 0.5, the test statistic $HCM_n$ seems robust against different bandwidths. Thus we choose the bandwidth $h=0.65$ in the simulations.

The empirical sizes and powers are presented in Table 3 and 4. We can observe that the results are similar to the case in Example 1 in the first three models. The proposed test $HCM_n$ still performs the best. It seems the unknown function $g(\cdot)$ in partial linear regression models does not impact the performance of the test. The situation becomes different in model $H_{24}$. When the dimension $q$ of the covariate $X$ is relatively large, all tests perform very bad. This can be explained that when $q$ is large, the weight of $T$ contributed to the test statistics become small.

\subsection{Real data analysis}
In this subsection we analyze two data sets for illustrations. The first one is the well-known baseball salary data set that can be obtain through the website \url{http://www4.stat.ncsu.edu/ ~ boos/var.select/baseball.html}. It contains contains 337 Major League Baseball players on the salary Y and 16 performance measures during both the 1991 and 1992 seasons. More descriptions of the variables in the salary data set can be found in the above website. Recently, Tan and Zhu (2018) analysed the data set and suggest to fit the data set by a parametric single-index model as following:
$$ Y=a+b (\beta^{\top}X)+c(\beta^{\top}X)^2 + \varepsilon. $$
Here we further investigate whether there exists a heteroscedasticity structure in the present model. We first plot the residuals $\hat{\varepsilon}$ against the fitted values $\hat{Y}$ in Figure 3, where $\hat{\varepsilon}=Y-\hat{a}-\hat{b}(\hat{\beta}^{\top}X)-\hat{c}(\hat{\beta}^{\top}X)^2$ and $\hat{Y}=\hat{a}+\hat{b}(\hat{\beta}^{\top}X)+\hat{c}(\hat{\beta}^{\top}X)^2$.
This plot shows that the heteroscedasticity structure may exist. When the proposed test is applied, we found the p-value is about 0. This indicates  the existence of heteroscedasticity. Thus a parametric single index model with heteroscedasticity is plausible for the salary data set.
$$ \rm  Figures \ 3 \ is \ about \ here $$

In the next example we consider the ACTG315 data set which is obtain from an AIDS clinical trial group study. This study tries to find the relationship between virologic and immunologic responses in AIDS clinical trials. The data set has been studied by Wu and Wu (2001, 2002) and Yang, Xue and Cheng (2009). Generally speaking, the virologic response RNA (measured by viral load) and immunologic response (measured by CD cell counts) have a negative correlation during the clinical trials. Let viral load be the response variable and CD4+cell counts and treatment time be the covariates variables. Liang et al. (2004) find that there is a linear relationship between viral load and CD4+ cell count, but a nonlinear relationship between viral load and treatment time. Base on this, Yang, Xue and Cheng (2009) suggested a partial linear regression model to fit the data. Xu and Guo (2013) further confirmed this by using a goodness of fit test. There are totally 317 observations available in the data set with 64 CD4+ cell counts missing. To illustrate our test, we clear the observations with missing variables. Let $Y$ be viral load, $T$ be treatment time and $X$ be CD4+cell counts. Yang, Xue and Cheng (2009) uses the following model for data fitting:
$$ Y=\beta X + g(T) +\varepsilon. $$
We further use the proposed test to check the existence of heteroscedasticity in the above models. The p-value is about 0.246. Thus we can not reject the homoscedasticity assumption in the partial linear regression model.
The scatter plot of the residuals $\hat{\varepsilon}$ against the fitted values $\hat{Y}$ is presented in Figure 4, where $\hat{\varepsilon}=Y-\hat{\beta}X-\hat{g}(T)$ and $\hat{Y}=\hat{\beta}X+\hat{g}(T)$. This plot also shows that a partial linear model with homoscedasticity is appropriate for the data set.
$$ \rm  Figures \ 4 \ is \ about \ here $$

\section{Conclusion and discussion}
In this paper we propose a test of heteroscedasticity by using a projection-based empirical process. Compared to existing tests of heteroscedasticity in the literature, the new test can detect the alternative hypotheses distinct from the null at a rate $O(1/\sqrt{n})$ that is the fastest convergence rate in hypothesis testing. It is also noted that we use all projected covariates $\alpha^{\top}Z$ to construct the test statistic and thus the test behaves as if the dimension of covariates was one. Therefore, the new test to some extent avoid the ``curse of dimensionality''. The simulation studies validate these theoretical results. The method can be easily extended to a more generalized problem of testing the parametric form of the variance function. But the limiting distributions of the empirical processes may have a more complicated structure which may lead the asymptotic test not available. This is beyond the scope of this paper and deserves a further study.

\section{Appendix.}
\subsection{Regularity Conditions}
In this subsection we give the regularity conditions for parametric regression models and partial linear regression models, respectively.  In the following, $C$ always stands for a constant that may be different from place to place.

First, we give the regularity conditions for parametric regression models~(\ref{3.1}) that are necessary to obtain the asymptotic properties of the test statistic. \\~\\
(A1) $E(\varepsilon^4) < \infty$; \\~\\
(A2) The parametric regression function $m(z, \beta)$ is twice continuously differentiable with respect to $\beta$ in a neighborhood $\Theta_0$. Let $m'(z, \beta)$ and $m''(z, \beta)$ be the first derivative and second derivative, respectively. Assume
$E\|m'(z, \beta)\|^2 < \infty$ and $E\|m''(z, \gamma)\| < \infty$ for any $\gamma \in \Theta_0$.

Next, we present the regularity conditions assumed for partial linear regression models~(\ref{3.2}). \\~\\
(B1) Let $E'(Y|T=t)$ be the derivative of $E(Y|T=t)$ and $F(x|t)$ be the conditional distribution function of $X$ given $T=t$. Suppose that
there exists a open neighborhood $\Theta_1$ of 0 such that for all $t$ and $x$,
\begin{eqnarray*}
&|E(X|T=t+u)-E(X|T=u)| \leq C|u|, \quad \forall \  u  \in \Theta_1; \\
&|E'(X|T=t+u)-E'(X|T=u)| \leq C|u|, \quad \forall \  u  \in \Theta_1; \\
&|F(x|t+u)-F(x|t)| \leq C|u|, \quad \forall \  u  \in \Theta_1.
\end{eqnarray*}
(B2) $E(Y^4) < \infty, E(\|X\|^4) < \infty$, and there exists a constant $C$ such that
$$ |E(\varepsilon^2| T=t, X=x)| \leq C, \quad \forall \ t \ {\rm and} \ x.  $$
(B3) The kernel function $K(\cdot)$ is bounded, continuous, symmetric about 0 and satisfies: (a) the support of $K(\cdot)$ is the interval $[-1,1]$;
(b) $\int_{-1}^1 K(u)du=1$ and $\int_{-1}^1|u|K(u)du \neq 0$.  \\~\\
(B4) $\sqrt{n}h^2 \to 0$ and $\sqrt{n} h \to \infty$, as $n \to \infty$.

The conditions (B1), (B2) and (B3) are common used in deriving the asymptotic properties of the nonparametric estimates. Condition (B4) is necessary to obtain the limiting distribution of the test statistics.

\subsection{Proofs of Theorems.}
\renewcommand{\theequation}{A.\arabic{equation}}
\setcounter{equation}{0}

{\bf Proof of Theorem 1.} Recall that $\hat{\eta}_i=\hat{\varepsilon}_i^2-\hat{\sigma}^2$ and $\hat{\varepsilon}_i=Y_i-m(Z_i, \hat{\beta})$. Then it follows that
$$\hat{\varepsilon}_i=\varepsilon_i-[m(Z_i, \hat{\beta})-m(Z_i, \beta)].$$
Consequently,
\begin{eqnarray*}
\hat{\eta}_i &=& \varepsilon_i^2+[m(Z_i, \hat{\beta})-m(Z_i, \beta)]^2-2\varepsilon_i[m(Z_i, \hat{\beta})-m(Z_i, \beta)]  \\
             &&  -\frac{1}{n} \sum_{j=1}^n \varepsilon_j^2 -\frac{1}{n}\sum_{j=1}^n[m(Z_j, \hat{\beta})-m(Z_j, \beta)]^2+\frac{2}{n}\sum_{j=1}^n \varepsilon_j[m(Z_j, \hat{\beta})-m(Z_j, \beta)]\\
             &=& \varepsilon_i^2-\sigma^2+\sigma^2-\frac{1}{n} \sum_{j=1}^n \varepsilon_j^2 \\
             &&  +[m(Z_i, \hat{\beta})-m(Z_i, \beta)]^2-\frac{1}{n}\sum_{j=1}^n[m(Z_j, \hat{\beta})-m(Z_j, \beta)]^2\\
             &&  -\{2\varepsilon_i[m(Z_i, \hat{\beta})-m(Z_i, \beta)]-\frac{2}{n}\sum_{j=1}^n \varepsilon_j[m(Z_j, \hat{\beta})-m(Z_j, \beta)] \}\\
             &=:& T_{1n}+T_{2n}-T_{3n}
\end{eqnarray*}
Let $V_{jn}(\alpha, t)= (1/\sqrt{n})\sum_{i=1}^n T_{jn} I(\alpha^{\top}Z_i \leq t)$. Then it follows that
$$V_n(\alpha, t)=V_{1n}(\alpha, t)+V_{2n}(\alpha, t)-V_{3n}(\alpha, t).$$
For $V_{1n}(\alpha, t)$, it is easy to see that
\begin{eqnarray*}
V_{1n}(\alpha, t) &=& \frac{1}{\sqrt{n}} \sum_{i=1}^n [\varepsilon_i^2-\sigma^2+ (\sigma^2-\frac{1}{n} \sum_{j=1}^n \varepsilon_j^2)]I(\alpha^{\top}Z_i \leq t) \\
&=& \frac{1}{\sqrt{n}} \sum_{i=1}^n \eta_i I(\alpha^{\top}Z_i \leq t) - \frac{1}{\sqrt{n}} \sum_{i=1}^n I(\alpha^{\top}Z_i \leq t) \frac{1}{n} \sum_{i=1}^n \eta_i.
\end{eqnarray*}
By Theorem 24 of Chapter 2 in Pollard (1984), we obtain that
$$\sup_{\alpha, t} |\frac{1}{n} \sum_{i=1}^n I(\alpha^{\top}Z_i \leq t)- F_{\alpha}(t)|=o_p(1).$$
Since $E(\varepsilon^4) < \infty$, it follows that
$$ \frac{1}{\sqrt n} \sum_{i=1}^n \eta_i= O_p(1). $$
Consequently,
$$ V_{1n}(\alpha, t)= \frac{1}{\sqrt{n}} \sum_{i=1}^n \eta_i [I(\alpha^{\top}Z_i \leq t)- F_{\alpha}(t)]+o_p(1), $$
uniformly in $(\alpha, t)$. To prove this theorem, it suffices to show that
$$ V_{2n}(\alpha, t)=o_p(1) \quad {\rm and } \quad  V_{3n}(\alpha, t)=o_p(1) \ {\rm uniformly \ in \ (\alpha, t)}. $$
For $ V_{2n}(\alpha, t)$, we have
\begin{eqnarray*}
V_{2n}(\alpha, t) &=& \frac{1}{\sqrt{n}} \sum_{i=1}^n \{[m(Z_i, \hat{\beta})-m(Z_i, \beta)]^2-\frac{1}{n}\sum_{j=1}^n[m(Z_j, \hat{\beta})-m(Z_j, \beta)]^2\} I(\alpha^{\top}Z_i \leq t)\\
&=&  \frac{1}{\sqrt{n}} \sum_{i=1}^n [m(Z_i, \hat{\beta})-m(Z_i, \beta)]^2 I(\alpha^{\top}Z_i \leq t) -\\
&&   \{ \frac{1}{\sqrt{n}} \sum_{i=1}^n I(\alpha^{\top}Z_i \leq t) \} \{ \frac{1}{n}\sum_{j=1}^n[m(Z_j, \hat{\beta})-m(Z_j, \beta)]^2 \} \\
&=:&  V_{21n}(\alpha, t)-V_{22n}(\alpha, t)
\end{eqnarray*}
By Taylor's expansion, we obtain
$$ m(Z_i, \hat{\beta})-m(Z_i, \beta)= (\hat \beta - \beta)^{\top}m'(Z_i, \beta) + \frac{1}{2}(\hat \beta - \beta)^{\top}m''(Z_i, \beta_1)(\hat \beta - \beta), $$
where $\beta_1$ lies between $\hat \beta$ and $\beta$. Then it follows that
\begin{eqnarray*}
V_{21n}(\alpha, t)&=&\frac{1}{\sqrt{n}} \sum_{i=1}^n (\hat \beta - \beta)^{\top}m'(Z_i, \beta)m'(Z_i, \beta)^{\top}(\hat \beta - \beta)I(\alpha^{\top}Z_i \leq t) + \\
&& \frac{1}{4\sqrt{n}} \sum_{i=1}^n [(\hat \beta - \beta)^{\top}m''(Z_i, \beta_1)(\hat \beta - \beta)]^2I(\alpha^{\top}Z_i \leq t) +\\
&& \frac{1}{\sqrt{n}} \sum_{i=1}^n (\hat \beta - \beta)^{\top}m'(Z_i, \beta) (\hat \beta - \beta)^{\top}m''(Z_i, \beta_1)(\hat \beta - \beta)I(\alpha^{\top}Z_i \leq t)
\end{eqnarray*}
Since $E\|m'(Z,\beta) \|^2 < \infty$ and $E\|m''(Z, \beta) \|^2 < \infty$ for all $\beta$, it is easy to see that
$$ V_{21n}(\alpha, t)= O_p(\frac{1}{\sqrt n}) \quad {\rm uniformly \ in \ (\alpha, t)}. $$
Similarly, we obtain that $ V_{22n}(\alpha, t)= O_p(1/\sqrt{n}) $ uniformly in $(\alpha, t)$.

Next we consider the third term $V_{3n}(\alpha, t)$ in $V_{n}(\alpha, t)$. Note that
\begin{eqnarray*}
V_{3n}(\alpha, t) &=& \frac{2}{\sqrt{n}} \sum_{i=1}^n \varepsilon_i [m(Z_i, \hat{\beta})-m(Z_i, \beta)] I(\alpha^{\top}Z_i \leq t)-\\
&& \{ \frac{2}{n} \sum_{i=1}^n \varepsilon_i [m(Z_i, \hat{\beta})-m(Z_i, \beta)]\} \{ \frac{1}{\sqrt{n}} \sum_{i=1}^n I(\alpha^{\top}Z_i \leq t)\} \\
&=:& V_{31n}(\alpha, t)-V_{32n}(\alpha, t).
\end{eqnarray*}
For $V_{31n}(\alpha, t)$, we have
\begin{eqnarray*}
V_{31n}(\alpha, t)&=&\frac{2}{\sqrt{n}} \sum_{i=1}^n \varepsilon_i (\hat \beta - \beta)^{\top}m'(Z_i, \beta) I(\alpha^{\top}Z_i \leq t)- \\
&& \frac{1}{\sqrt{n}} \sum_{i=1}^n \varepsilon_i (\hat \beta - \beta)^{\top}m''(Z_i, \beta_1)(\hat \beta - \beta) I(\alpha^{\top}Z_i \leq t)\\
&=& o_p(1)
\end{eqnarray*}
By a similar argument, we obtain $V_{32n}(\alpha, t)=o_p(1)$ uniformly in $(\alpha, t)$. Altogether we complete the proof. \hfill$\Box$

{\bf Proof of Theorem 2.} First we give the proof for the test statistic $C_n$. Let
$$V_n(\alpha, t)=\frac{1}{\sqrt{n}} \sum_{i=1}^n \hat{\eta}_i I(\alpha^{\top}Z_i \leq t).$$
Here $\hat{\eta}_i=\hat{\varepsilon}_i^2-\hat{\sigma}^2$, $\hat{\varepsilon}_i=Y_i-\hat{\beta}^{\top}X_i-\hat{g}(T_i)$, and $\hat{\sigma}^2=(1/n)\sum_{i=1}^n \hat{\varepsilon}_i^2$. Recall that
$$ \hat{g}(T_i)=\frac{1}{n} \sum_{j \neq i}^n [Y_j-\hat{\beta}^{\top}X_j]K_h(T_i-T_j)/\hat{f}_i(T_i). $$
Then it can be decomposed as following
$$ \hat{g}(T_i)=\tilde{g}(T_i)-(\hat{\beta}-\beta)^{\top} \frac{1}{n} \sum_{j \neq i}^n X_j K_h(T_i-T_j)/\hat{f}_i(T_i), $$
where $\tilde{g}(T_i)=(1/n) \sum_{j \neq i}^n [Y_j-\beta^{\top}X_j]K_h(T_i-T_j)/\hat{f}_i(T_i)$. Consequently, we obtain that
$$ \hat{\varepsilon}_i=\varepsilon_i- (\hat{\beta}-\beta)^{\top} \{X_i - \hat{E}(X|T_i)\}- \{\tilde{g}(T_i)-g(T_i)\}.$$
Then it follows that
\begin{eqnarray*}
\hat{\eta}_i&=&\varepsilon_i^2-\sigma^2+ (\sigma^2-\frac{1}{n} \sum_{i=1}^n \varepsilon_i^2) \\
&& + \{(\hat{\beta}-\beta)^{\top} [X_i - \hat{E}(X|T_i)]\}^2-\frac{1}{n}\sum_{i=1}^n \{(\hat{\beta}-\beta)^{\top} [X_i - \hat{E}(X|T_i)]\}^2 \\
&& + [\tilde{g}(T_i)-g(T_i)]^2 - \frac{1}{n} \sum_{i=1}^n [\tilde{g}(T_i)-g(T_i)]^2 \\
&& - 2\varepsilon_i (\hat{\beta}-\beta)^{\top} [X_i - \hat{E}(X|T_i)] + (\hat{\beta}-\beta)^{\top} \frac{2}{n} \sum_{i=1}^n \varepsilon_i[X_i - \hat{E}(X|T_i)]\\
&& - 2\varepsilon_i [\tilde{g}(T_i)-g(T_i)] + \frac{2}{n} \sum_{i=1}^n \varepsilon_i [\tilde{g}(T_i)-g(T_i)] \\
&& + 2(\hat{\beta}-\beta)^{\top} [X_i - \hat{E}(X|T_i)][\tilde{g}(T_i)-g(T_i)]-(\hat{\beta}-\beta)^{\top} \frac{2}{n} \sum_{i=1}^n [X_i - \hat{E}(X|T_i)][\tilde{g}(T_i)-g(T_i)]\\
&=:& T_{1n}+T_{2n}+T_{3n}-T_{4n}-T_{5n}+T_{6n}.
\end{eqnarray*}
Let $V_{jn}(\alpha, t)= (1/\sqrt{n})\sum_{i=1}^n T_{jn} I(\alpha^{\top}Z_i \leq t)$. First we consider $V_{1n}(\alpha, t)$. Note that
\begin{eqnarray*}
V_{1n}(\alpha, t) &=& \frac{1}{\sqrt{n}} \sum_{i=1}^n [\varepsilon_i^2-\sigma^2+ (\sigma^2-\frac{1}{n} \sum_{i=1}^n \varepsilon_i^2)]I(\alpha^{\top}Z_i \leq t) \\
&=& \frac{1}{\sqrt{n}} \sum_{i=1}^n \eta_i I(\alpha^{\top}Z_i \leq t) - \frac{1}{\sqrt{n}} \sum_{i=1}^n I(\alpha^{\top}Z_i \leq t) \frac{1}{n} \sum_{i=1}^n \eta_i.
\end{eqnarray*}
By standard empirical process theory, see, e.g. Pollard (1984, Chapter II), we have
$$\sup_{\alpha, t} |\frac{1}{n} \sum_{i=1}^n I(\alpha^{\top}Z_i \leq t)- F_{\alpha}(t)|=o_p(1).$$
Then it follows that
$$ V_{1n}(\alpha, t)= \frac{1}{\sqrt{n}} \sum_{i=1}^n \eta_i [I(\alpha^{\top}Z_i \leq t)- F_{\alpha}(t)]+o_p(1), $$
uniformly in $(\alpha, t)$. Next we show that the rest terms $V_{jn}(\alpha, t) = o_p(1)$ uniformly in $(\alpha, t)$ for $2 \leq j \leq 6$.

Recall that
\begin{eqnarray*}
V_{2n}(\alpha, t)&=&\frac{1}{\sqrt{n}} \sum_{i=1}^n  \{(\hat{\beta}-\beta)^{\top} [X_i - \hat{E}(X|T_i)]\}^2 I(\alpha^{\top}Z_i \leq t)-\\
&&\frac{1}{n}\sum_{i=1}^n \{(\hat{\beta}-\beta)^{\top} [X_i - \hat{E}(X|T_i)]\}^2  \frac{1}{\sqrt{n}} \sum_{i=1}^n I(\alpha^{\top}Z_i \leq t).
\end{eqnarray*}
Then it follows that
\begin{eqnarray*}
V_{2n}(\alpha, t)&=&\frac{1}{\sqrt{n}} \sum_{i=1}^n  \{(\hat{\beta}-\beta)^{\top} [X_i - E(X|T_i)]\}^2 I(\alpha^{\top}Z_i \leq t)-\\
&&\frac{1}{n}\sum_{i=1}^n \{(\hat{\beta}-\beta)^{\top} [X_i - E(X|T_i)]\}^2  \frac{1}{\sqrt{n}} \sum_{i=1}^n I(\alpha^{\top}Z_i \leq t)+\\
&&\frac{1}{\sqrt{n}} \sum_{i=1}^n  \{(\hat{\beta}-\beta)^{\top} [\hat{E}(X|T_i) - E(X|T_i)]\}^2 I(\alpha^{\top}Z_i \leq t)-\\
&&\frac{1}{n}\sum_{i=1}^n \{(\hat{\beta}-\beta)^{\top} [\hat{E}(X|T_i) - E(X|T_i)]\}^2  \frac{1}{\sqrt{n}} \sum_{i=1}^n I(\alpha^{\top}Z_i \leq t)-\\
&&\frac{2}{\sqrt{n}} \sum_{i=1}^n (\hat{\beta}-\beta)^{\top} [X_i - E(X|T_i)](\hat{\beta}-\beta)^{\top} [\hat{E}(X|T_i) - E(X|T_i)]I(\alpha^{\top}Z_i \leq t)+\\
&&\frac{2}{n} \sum_{i=1}^n (\hat{\beta}-\beta)^{\top} [X_i - E(X|T_i)](\hat{\beta}-\beta)^{\top} [\hat{E}(X|T_i) - E(X|T_i)] \frac{1}{\sqrt{n}}\sum_{i=1}^n I(\alpha^{\top}Z_i \leq t).
\end{eqnarray*}
Since $\sup_{t}|\hat{E}(X|T=t) - E(X|T=t)|=O_p(\log{n}/\sqrt{nh}+h)$, we obtain that
$$ V_{2n}(\alpha, t)=O_p(\frac{1}{\sqrt{n}})+ O_p(\frac{1}{\sqrt{n}})O_p(\frac{\log{n}}{\sqrt{nh}}+h)^2+O_p(\frac{1}{\sqrt{n}})O_p(\frac{\log{n}}{\sqrt{nh}}+h). $$
Thus we have $ V_{2n}(\alpha, t)=o_p(1)$ uniformly in $(\alpha, t)$. By similar arguments, we can show that $ V_{jn}(\alpha, t)=o_p(1)$ uniformly in  $(\alpha, t)$ for $j=3, 4, 6$.

Now we consider the term $ V_{5n}(\alpha, t)$. Note that
\begin{eqnarray*}
V_{5n}(\alpha, t)&=&\frac{2}{\sqrt{n}} \sum_{i=1}^n \varepsilon_i [\tilde{g}(T_i)-g(T_i)]I(\alpha^{\top}Z_i \leq t) -
    \frac{2}{\sqrt{n}} \sum_{i=1}^n I(\alpha^{\top}Z_i \leq t) \frac{1}{n} \sum_{i=1}^n \varepsilon_i [\tilde{g}(T_i)-g(T_i)] \\
&=& \frac{2}{\sqrt{n}} \sum_{i=1}^n \varepsilon_i [\tilde{g}(T_i)-g(T_i)] [I(\alpha^{\top}Z_i \leq t)-F_{\alpha}(t)] - \\
&&  \frac{2}{\sqrt{n}} \sum_{i=1}^n [I(\alpha^{\top}Z_i \leq t)-F_{\alpha}(t)] \frac{1}{n} \sum_{i=1}^n \varepsilon_i [\tilde{g}(T_i)-g(T_i)]
\end{eqnarray*}
Then it follows that
$$ V_{5n}(\alpha, t)=\frac{2}{\sqrt{n}} \sum_{i=1}^n \varepsilon_i [\tilde{g}(T_i)-g(T_i)] [I(\alpha^{\top}Z_i \leq t)-F_{\alpha}(t)] + O_p(\log{n}/{\sqrt{nh}}+h),$$
uniformly in $(\alpha, t)$. Let $r(t)=g(t)f(t)$ and $ \hat{r}(T_i)=(1/n) \sum_{j \neq i}^n [Y_j-\beta^{\top}X_j]K_h(T_i-T_j)$. Then $\hat{r}(T_i)=\tilde{g}(T_i)\hat{f}(T_i)$. Consequently,
\begin{eqnarray*}
\tilde{g}(T_i)-g(T_i)&=&\frac{\hat{r}(T_i)}{\hat{f}(T_i)} - \frac{r(T_i)}{f(T_i)}=\frac{\hat{r}(T_i)-r(T_i)}{f(T_i)}
-g(T_i)\frac{\hat{f}(T_i)-f(T_i)}{f(T_i)} \\
&& -\frac{[\hat{r}(T_i)-r(T_i)][\hat{f}(T_i)-f(T_i)]}{\hat{f}(T_i)f(T_i)}+\frac{g(T_i)[\hat{f}(T_i)-f(T_i)]^2}{\hat{f}(T_i)f(T_i)}.
\end{eqnarray*}
Then we obtain that
\begin{eqnarray*}
V_{5n}(\alpha, t)&=&\frac{2}{\sqrt{n}} \sum_{i=1}^n \varepsilon_i \frac{\hat{r}(T_i)-r(T_i)}{f(T_i)} [I(\alpha^{\top}Z_i \leq t)-F_{\alpha}(t)] - \\
&&  \frac{2}{\sqrt{n}} \sum_{i=1}^n \varepsilon_i g(T_i)\frac{\hat{f}(T_i)-f(T_i)}{f(T_i)} [I(\alpha^{\top}Z_i \leq t)-F_{\alpha}(t)] + \\
&&  O_p(\log{n}/{\sqrt{nh}}+h) + O_p(\sqrt{n}) O_p(\log{n}/{\sqrt{nh}}+h)^2 \\
&=:& J_{1n}+J_{2n} + O_p(\log{n}/{\sqrt{nh}}+h) + O_p((\log{n})^2/{\sqrt{n}h}+ \sqrt{n} h^2)
\end{eqnarray*}
It will be shown that $J_{1n}$ and $J_{2n}$ converge to zero in probability uniformly in $(\alpha, t)$. We only give the detailed arguments for $J_{1n}$. The arguments for $J_{2n}$ are similar. Note that
\begin{eqnarray*}
J_{1n}&=& \frac{2}{\sqrt{n}} \sum_{i=1}^n \frac{\varepsilon_i}{f(T_i)} \left(\frac{1}{nh} \sum_{j \neq i}^n (Y_j-\beta^{\top}X_j)K(\frac{T_i-T_j}{h})-r(T_i)\right) [I(\alpha^{\top}Z_i \leq t)-F_{\alpha}(t)] \\
&=& \frac{2}{hn^{3/2}} \sum_{i \neq j}^n \frac{\varepsilon_i}{f(T_i)} \left\{[g(T_j)+\varepsilon_j]K(\frac{T_i-T_j}{h})- hr(T_i) \right\} [I(\alpha^{\top}Z_i \leq t)-F_{\alpha}(t)].
\end{eqnarray*}
Define $\tau_i=(X_i, T_i, \varepsilon_i)$ and
$$ f_{\alpha, t}(\tau_i, \tau_j)= \frac{\varepsilon_i}{f(T_i)} \left\{[g(T_j)+\varepsilon_j]K(\frac{T_i-T_j}{h})- hr(T_i) \right\} [I(\alpha^{\top}Z_i \leq t)-F_{\alpha}(t)].$$
Then it follows that
$$J_{1n}=\frac{2}{hn^{3/2}} \sum_{i \neq j}^n f_{\alpha, t}(\tau_i, \tau_j)=\frac{1}{hn^{3/2}} \sum_{i \neq j}^n [f_{\alpha, t}(\tau_i, \tau_j)+f_{\alpha, t}(\tau_j, \tau_i)]$$
Let $w_{\alpha, t}(\tau_i, \tau_j)=f_{\alpha, t}(\tau_i, \tau_j)+f_{\alpha, t}(\tau_j, \tau_i)$ and define
$$ \tilde{J}_{1n}= \sum_{i \neq j}^n \{w_{\alpha, t}(\tau_i, \tau_j)-E(w_{\alpha, t}(\tau_i, \tau_j)|\tau_i)-E(w_{\alpha, t}(\tau_i, \tau_j)|\tau_j)  \}. $$
Then $\tilde{J}_{1n}$ is a $\mathbb{P}$-degenerate $U$-process (see, Nolan and Pollard (1987)). Here $\mathbb{P}$ is the probability measure of $(X, T, \varepsilon)$. Consider the class of functions
$$ \mathcal{F}_n=\{w_{\alpha, t}(\tau_1, \tau_2)-E(w_{\alpha, t}(\tau_1, \tau_2)|\tau_1)-E(w_{\alpha, t}(\tau_1, \tau_2)|\tau_2): \alpha \in \mathbb{S}^{p+1}, \ t \in \mathbb{R} \}.  $$
Then $\mathcal{F}_n$ is a $\mathbb{P}$-degenerate class of functions with an envelope
\begin{eqnarray*}
G_n(\tau_1, \tau_2)&=&|\frac{\varepsilon_1}{f(T_1)} \left\{ [g(T_2)+\varepsilon_2]K(\frac{T_1-T_2}{h}) - \int g(t)f(t)K(\frac{T_1-t}{h})dt \right\}|+ \\
&& | \frac{\varepsilon_2}{f(T_2)} \left\{ [g(T_1)+\varepsilon_1]K(\frac{T_2-T_1}{h}) - \int g(t)f(t)K(\frac{T_2-t}{h})dt \right\} |
\end{eqnarray*}
It is well known that the class of indictor functions is a VC-class. Then it follows that
$$ N_2\{ u(T_n G_n^2)^{1/2}, L_2(T_n), \mathcal{F}_n\} \leq C u^{-w}, $$
where the constants $C$ and $w$ do not depend on $n$,
$$ T_n g^2= \sum_{i \neq j} [g^2(\tau_{2i}, \tau_{2j}) + g^2(\tau_{2i}, \tau_{2j-1}) +g^2(\tau_{2i-1}, \tau_{2j}) +g^2(\tau_{2i-1}, \tau_{2j-1})], $$
and $N_2\{u, L_2(T_n), \mathcal{F}_n\}$ is the covering number of $\mathcal{F}_n$ under the semi-metric $L_2(T_n)$.
By Theorem 6 of Nolan and Pollard (1987), we obtain that
$$ E (\sup_{\alpha, t} |\tilde{J}_{1n}| ) \leq C E[\theta_n + \gamma_n J_n(\theta_n/\gamma_n)], $$
where $C$ is a universal constant, $\gamma_n=(T_n G_n^2)^{1/2}$, $\theta_n=(1/4)\sup_{\mathcal{F}_n} (T_ng^2)^{1/2}$, and
$$J_n(s)= \int_0^s \log N_2\{ u(T_n G_n^2)^{1/2}, L_2(T_n), \mathcal{F}_n\} du.$$
Therefore, we have
$$  E(\sup_{\alpha, t} |\tilde{J}_{1n}|) \leq C E[\gamma_n/4+J_n(1/4)\gamma_n] \leq C E(T_n G_n^2)^{1/2}.$$
It is easy to see that $E(T_n G_n^2)=O(n^2h)$. Thus we obtain that $\sup_{\alpha, t} |\tilde{J}_{1n}|=O_p(nh^{1/2})$.

Recall that
$$ J_{1n}= \frac{1}{hn^{3/2}} \tilde{J}_{1n} + \frac{1}{hn^{3/2}} \sum_{i \neq j}^n \{ E(w_{\alpha, t}(\tau_i, \tau_j)|\tau_i)+E(w_{\alpha, t}(\tau_i, \tau_j)|\tau_j) \}. $$
To prove $J_{1n}=o_p(1)$ uniformly in $(\alpha, t)$, it remains to show that
$$ \frac{1}{hn^{3/2}} \sum_{i \neq j}^n \{ E(w_{\alpha, t}(\tau_i, \tau_j)|\tau_i)+E(w_{\alpha, t}(\tau_i, \tau_j)|\tau_j) \}=o_p(1), $$
uniformly in  $(\alpha, t)$. Note that
\begin{eqnarray*}
&& \frac{1}{hn^{3/2}} \sum_{i \neq j}^n E(w_{\alpha, t}(\tau_i, \tau_j)|\tau_i) \\
&=&\frac{n-1}{hn^{3/2}}\sum_{i=1}^n \frac{\varepsilon_i}{f(T_i)}[\int r(t) K(\frac{T_i-t}{h})dt-hr(T_i)][I(\alpha^{\top}Z_i \leq t)-F_{\alpha}(t)] \\
&=& \frac{n-1}{n} \frac{1}{\sqrt{n}} \sum_{i=1}^n \frac{\varepsilon_i}{f(T_i)} [\int r(T_i-hu)K(u)du-r(T_i)] [I(\alpha^{\top}Z_i \leq t)-F_{\alpha}(t)] \\
&=& \frac{n-1}{n} \frac{1}{\sqrt{n}} \sum_{i=1}^n \frac{\varepsilon_i}{f(T_i)} [\int (-hu)r'(\zeta_i)K(u)du] [I(\alpha^{\top}Z_i \leq t)-F_{\alpha}(t)] \\
&=& O_p(h),
\end{eqnarray*}
where $\zeta_i$ lies between $T_i$ and $T_i- hu$. Similarly, we have $(1/hn^{3/2}) \sum_{i \neq j}^n E(w_{\alpha, t}(\tau_i, \tau_j)|\tau_j)=O_p(h)$ uniformly in $(\alpha, t)$. Thus $J_{1n}=o_p(1)$ uniformly in $(\alpha, t)$. Altogether, we obtain
$$ V_{n}(\alpha, t)= \frac{1}{\sqrt{n}} \sum_{i=1}^n \eta_i [I(\alpha^{\top}Z_i \leq t)- F_{\alpha}(t)]+o_p(1), $$
Hence we complete the proof.   \hfill$\Box$

{\bf Proof of Theorem 3.} Similar to the arguments in Theorem 1, we have
$$ V_{n}(\alpha, t)= \frac{1}{\sqrt{n}} \sum_{i=1}^n (\varepsilon_i^2-\sigma^2) [I(\alpha^{\top}Z_i \leq t)- F_{\alpha}(t)]+o_p(1). $$
Then under the alternative $H_{1n}$, we obtain that
\begin{eqnarray*}
V_{n}(\alpha, t) &=& \frac{1}{\sqrt{n}} \sum_{i=1}^n [\varepsilon_i^2-\sigma^2-c_n s(Z_i)] [I(\alpha^{\top}Z_i \leq t)- F_{\alpha}(t)]+ \\
&&  c_n \frac{1}{\sqrt{n}} \sum_{i=1}^n s(Z_i) [I(\alpha^{\top}Z_i \leq t)- F_{\alpha}(t)]+  o_p(1).
\end{eqnarray*}
If $c_n= 1/\sqrt{n}$, then the first sum in $V_{n}(\alpha, t)$ converges to $V_{\infty}(\alpha, t)$ and the seconds tends to $E\{s(Z) [I(\alpha^{\top}Z \leq t)- F_{\alpha}(t)]\}$. If $\sqrt{n}c_n \to \infty$, then $V_{n}(\alpha, t)$ tends to infinity. Altogether we complete the proof.       \hfill$\Box$

{\bf Proof of Theorem 4.} The proof of Theorem 4 follows the same line as in the proof of Theorem 3.2 in Zhu, Fujikoshi and Naito (2001) with some extra complications that arise from the indicator functions $I(\beta^{\top}Z \leq t)$ involving the projections. Since the class of indicator functions $$\mathcal{F}=\{f(z)=I(\beta^{\top}z \leq t): \beta \in \mathbb{S}^{p}, t \in \mathbb{R} \}$$
is also a VC-class, the proof can be very similar to the proof of Theorem 3.2 in Zhu, Fujikoshi and Naito (2001). Thus we omit it here.    \hfill$\Box$

\newpage
\begin{table}[ht!]\caption{Empirical sizes and powers of $HCM_n$, $T_n^{G}$, $T_n^{ZH}$, and $T_n^{ZFN}$ for $H_{11}$ and $H_{12}$ in Example 1.}
\centering
{\small\scriptsize\hspace{12.5cm}
\renewcommand{\arraystretch}{1}\tabcolsep 0.25cm
\begin{tabular}{cccccccccccc}
\hline
&\multicolumn{1}{c}{a} &\multicolumn{2}{c}{$HCM_n$} &\multicolumn{2}{c}{$T_n^{G}$} &\multicolumn{2}{c}{$T_n^{ZH}$} &\multicolumn{2}{c}{$T_n^{ZFN}$}  \\
&&\multicolumn{1}{c}{n=100}&\multicolumn{1}{c}{n=200} &\multicolumn{1}{c}{n=100}&\multicolumn{1}{c}{n=200} &\multicolumn{1}{c}{n=100}&\multicolumn{1}{c}{n=200} &\multicolumn{1}{c}{n=100}&\multicolumn{1}{c}{n=200}\\
\hline
$H_{11}, p=2$        &0.0       &0.045 &0.051   &0.058 &0.062   &0.042 &0.033   &0.052 &0.049\\
                     &0.1       &0.528 &0.895   &0.391 &0.751   &0.123 &0.286   &0.503 &0.889\\
                     &0.2       &0.966 &1.000   &0.921 &1.000   &0.468 &0.889   &0.961 &1.000\\
                     &0.3       &0.998 &1.000   &0.990 &1.000   &0.779 &0.990   &0.985 &1.000\\
                     &0.4       &0.998 &1.000   &0.999 &1.000   &0.885 &0.998   &0.974 &1.000\\
                     &0.5       &0.994 &1.000   &0.999 &1.000   &0.928 &1.000   &0.965 &0.998\\
\hline
$H_{11}, p=4$        &0.0           &0.055 &0.053   &0.050 &0.057   &0.031 &0.022   &0.063 &0.051\\
                     &0.1           &0.398 &0.767   &0.233 &0.481   &0.049 &0.095   &0.131 &0.593\\
                     &0.2           &0.874 &0.997   &0.669 &0.958   &0.145 &0.347   &0.426 &0.956\\
                     &0.3           &0.963 &1.000   &0.857 &0.999   &0.306 &0.621   &0.541 &0.964\\
                     &0.4           &0.970 &0.999   &0.943 &1.000   &0.430 &0.821   &0.419 &0.916\\
                     &0.5           &0.944 &0.998   &0.958 &1.000   &0.492 &0.876   &0.297 &0.809\\
\hline
$H_{11}, p=8$        &0.0       &0.049 &0.049   &0.053 &0.065   &0.045 &0.036   &0.050 &0.049\\
                     &0.1       &0.289 &0.600   &0.151 &0.257   &0.055 &0.055   &0.004 &0.004\\
                     &0.2       &0.755 &0.980   &0.352 &0.688   &0.108 &0.132   &0.004 &0.010\\
                     &0.3       &0.883 &0.997   &0.526 &0.892   &0.138 &0.187   &0.004 &0.010\\
                     &0.4       &0.874 &0.990   &0.623 &0.946   &0.167 &0.254   &0.009 &0.009\\
                     &0.5       &0.853 &0.988   &0.647 &0.966   &0.247 &0.324   &0.023 &0.014\\
\hline
$H_{12}, p=2$        &0.0           &0.054 &0.046   &0.043 &0.068   &0.032 &0.056   &0.052 &0.045\\
                     &0.1           &0.183 &0.347   &0.138 &0.262   &0.059 &0.080   &0.153 &0.327\\
                     &0.2           &0.564 &0.892   &0.440 &0.753   &0.121 &0.295   &0.502 &0.878\\
                     &0.3           &0.882 &0.996   &0.747 &0.967   &0.281 &0.692   &0.810 &0.993\\
                     &0.4           &0.973 &0.999   &0.927 &0.999   &0.514 &0.900   &0.919 &0.997\\
                     &0.5           &0.987 &0.999   &0.983 &1.000   &0.650 &0.964   &0.944 &0.986\\
\hline
$H_{12}, p=4$        &0.0               &0.050 &0.046   &0.058 &0.048   &0.028 &0.023   &0.057 &0.056\\
                     &0.1               &0.127 &0.270   &0.103 &0.157   &0.034 &0.038   &0.040 &0.110\\
                     &0.2               &0.424 &0.789   &0.264 &0.479   &0.048 &0.075   &0.104 &0.529\\
                     &0.3               &0.702 &0.976   &0.488 &0.856   &0.114 &0.208   &0.210 &0.804\\
                     &0.4               &0.862 &0.993   &0.727 &0.976   &0.163 &0.436   &0.294 &0.857\\
                     &0.5               &0.910 &0.993   &0.849 &0.996   &0.272 &0.651   &0.317 &0.802\\
\hline
$H_{12}, p=8$        &0.0           &0.050 &0.046   &0.085 &0.062   &0.039 &0.037   &0.054 &0.047\\
                     &0.1           &0.112 &0.193   &0.083 &0.111   &0.055 &0.053   &0.014 &0.001\\
                     &0.2           &0.274 &0.618   &0.156 &0.266   &0.063 &0.057   &0.002 &0.003\\
                     &0.3           &0.549 &0.919   &0.252 &0.526   &0.089 &0.086   &0.002 &0.000\\
                     &0.4           &0.757 &0.973   &0.372 &0.727   &0.113 &0.154   &0.001 &0.002\\
                     &0.5           &0.836 &0.972   &0.494 &0.865   &0.140 &0.207   &0.001 &0.002\\
\hline
\end{tabular}
}
\end{table}

\begin{table}[ht!]\caption{Empirical sizes and powers of $HCM_n$, $T_n^{G}$, $T_n^{ZH}$, and $T_n^{ZFN}$ for $H_{13}$ and $H_{14}$ in Example 1.}
\centering
{\small\scriptsize\hspace{12.5cm}
\renewcommand{\arraystretch}{1}\tabcolsep 0.25cm
\begin{tabular}{cccccccccccc}
\hline
&\multicolumn{1}{c}{a} &\multicolumn{2}{c}{$HCM_n$} &\multicolumn{2}{c}{$T_n^{G}$} &\multicolumn{2}{c}{$T_n^{ZH}$} &\multicolumn{2}{c}{$T_n^{ZFN}$}  \\
&&\multicolumn{1}{c}{n=100}&\multicolumn{1}{c}{n=200} &\multicolumn{1}{c}{n=100}&\multicolumn{1}{c}{n=200} &\multicolumn{1}{c}{n=100}&\multicolumn{1}{c}{n=200} &\multicolumn{1}{c}{n=100}&\multicolumn{1}{c}{n=200}\\
\hline
$H_{13}, p=2$      &0.0             &0.049 &0.050   &0.042 &0.054   &0.026 &0.043   &0.067 &0.048\\
                   &0.1             &0.102 &0.169   &0.110 &0.147   &0.036 &0.060   &0.081 &0.178\\
                   &0.2             &0.307 &0.555   &0.256 &0.488   &0.077 &0.187   &0.238 &0.576\\
                   &0.3             &0.566 &0.902   &0.467 &0.836   &0.184 &0.412   &0.516 &0.890\\
                   &0.4             &0.782 &0.993   &0.712 &0.977   &0.310 &0.687   &0.772 &0.986\\
                   &0.5             &0.922 &0.999   &0.892 &0.998   &0.497 &0.880   &0.910 &1.000\\
\hline
$H_{13}, p=4$      &0.0             &0.052 &0.057   &0.070 &0.060   &0.020 &0.030   &0.063 &0.055\\
                   &0.1             &0.089 &0.114   &0.107 &0.110   &0.038 &0.037   &0.022 &0.056\\
                   &0.2             &0.191 &0.406   &0.169 &0.288   &0.057 &0.067   &0.054 &0.219\\
                   &0.3             &0.389 &0.758   &0.295 &0.589   &0.060 &0.125   &0.125 &0.528\\
                   &0.4             &0.596 &0.931   &0.471 &0.834   &0.076 &0.215   &0.240 &0.816\\
                   &0.5             &0.756 &0.989   &0.635 &0.958   &0.138 &0.362   &0.344 &0.932\\
\hline
$H_{13}, p=8$      &0.0             &0.056 &0.043   &0.077 &0.071   &0.055 &0.044   &0.061 &0.052\\
                   &0.1             &0.075 &0.081   &0.079 &0.077   &0.054 &0.053   &0.022 &0.030\\
                   &0.2             &0.138 &0.266   &0.096 &0.144   &0.045 &0.064   &0.020 &0.011\\
                   &0.3             &0.261 &0.556   &0.149 &0.300   &0.053 &0.062   &0.008 &0.010\\
                   &0.4             &0.427 &0.833   &0.226 &0.445   &0.068 &0.072   &0.013 &0.011\\
                   &0.5             &0.602 &0.945   &0.308 &0.626   &0.075 &0.105   &0.010 &0.019\\
\hline
$H_{14}, p=2$      &0.0         &0.046 &0.048   &0.074 &0.064   &0.033 &0.036   &0.051 &0.052\\
                   &0.1         &0.582 &0.907   &0.421 &0.756   &0.139 &0.312   &0.596 &0.931\\
                   &0.2         &0.956 &0.999   &0.926 &1.000   &0.473 &0.883   &0.954 &1.000\\
                   &0.3         &0.991 &1.000   &0.997 &1.000   &0.783 &0.994   &0.989 &1.000\\
                   &0.4         &0.993 &1.000   &1.000 &1.000   &0.882 &0.999   &0.984 &1.000\\
                   &0.5         &0.992 &1.000   &1.000 &1.000   &0.927 &0.999   &0.975 &1.000\\
\hline
$H_{14}, p=4$      &0.0           &0.033 &0.053   &0.072 &0.055   &0.027 &0.029   &0.041 &0.048\\
                   &0.1           &0.448 &0.805   &0.281 &0.520   &0.060 &0.097   &0.319 &0.762\\
                   &0.2           &0.886 &0.998   &0.706 &0.974   &0.199 &0.366   &0.690 &0.983\\
                   &0.3           &0.945 &1.000   &0.885 &1.000   &0.292 &0.652   &0.718 &0.979\\
                   &0.4           &0.966 &0.998   &0.964 &1.000   &0.446 &0.805   &0.678 &0.961\\
                   &0.5           &0.966 &1.000   &0.963 &1.000   &0.504 &0.863   &0.560 &0.916\\
\hline
$H_{14}, p=8$      &0.0           &0.041 &0.042   &0.079 &0.059   &0.045 &0.045   &0.031 &0.047\\
                   &0.1           &0.332 &0.655   &0.164 &0.253   &0.072 &0.065   &0.010 &0.029\\
                   &0.2           &0.683 &0.972   &0.346 &0.698   &0.133 &0.139   &0.012 &0.054\\
                   &0.3           &0.838 &0.988   &0.538 &0.918   &0.166 &0.211   &0.003 &0.041\\
                   &0.4           &0.877 &0.992   &0.631 &0.959   &0.231 &0.273   &0.003 &0.035\\
                   &0.5           &0.882 &0.986   &0.677 &0.975   &0.278 &0.324   &0.003 &0.011\\
\hline
\end{tabular}
}
\end{table}

\begin{figure}
  \centering
  \includegraphics[width=14cm,height=8cm]{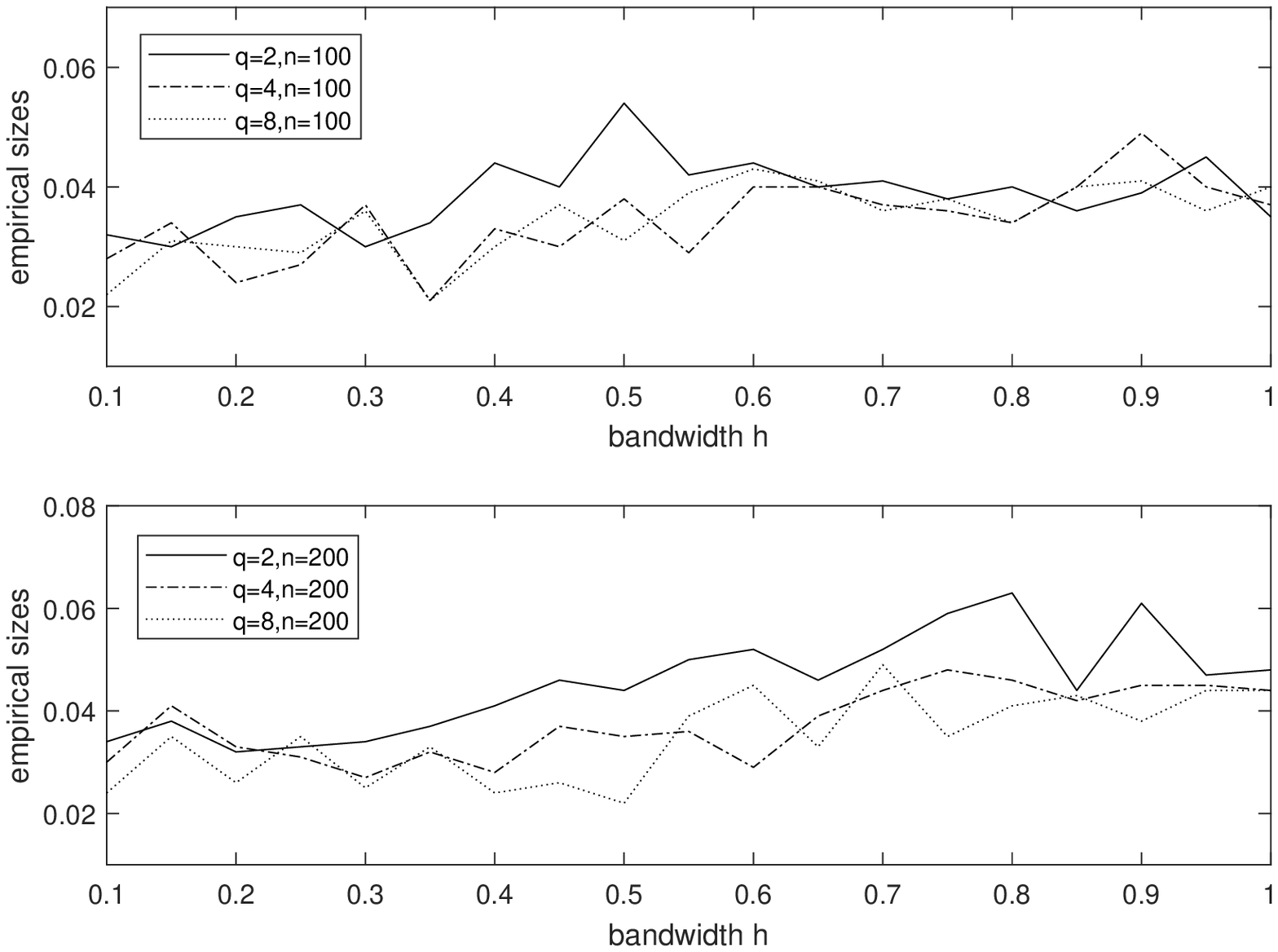}
  \caption{The empirical size curves of $HCM_n$ against the different bandwidths and sample size 100 and 200 with $a=0$ in Model~$H_{21}$.}\label{Figure 1}
\end{figure}

\begin{figure}
  \centering
  \includegraphics[width=14cm,height=8cm]{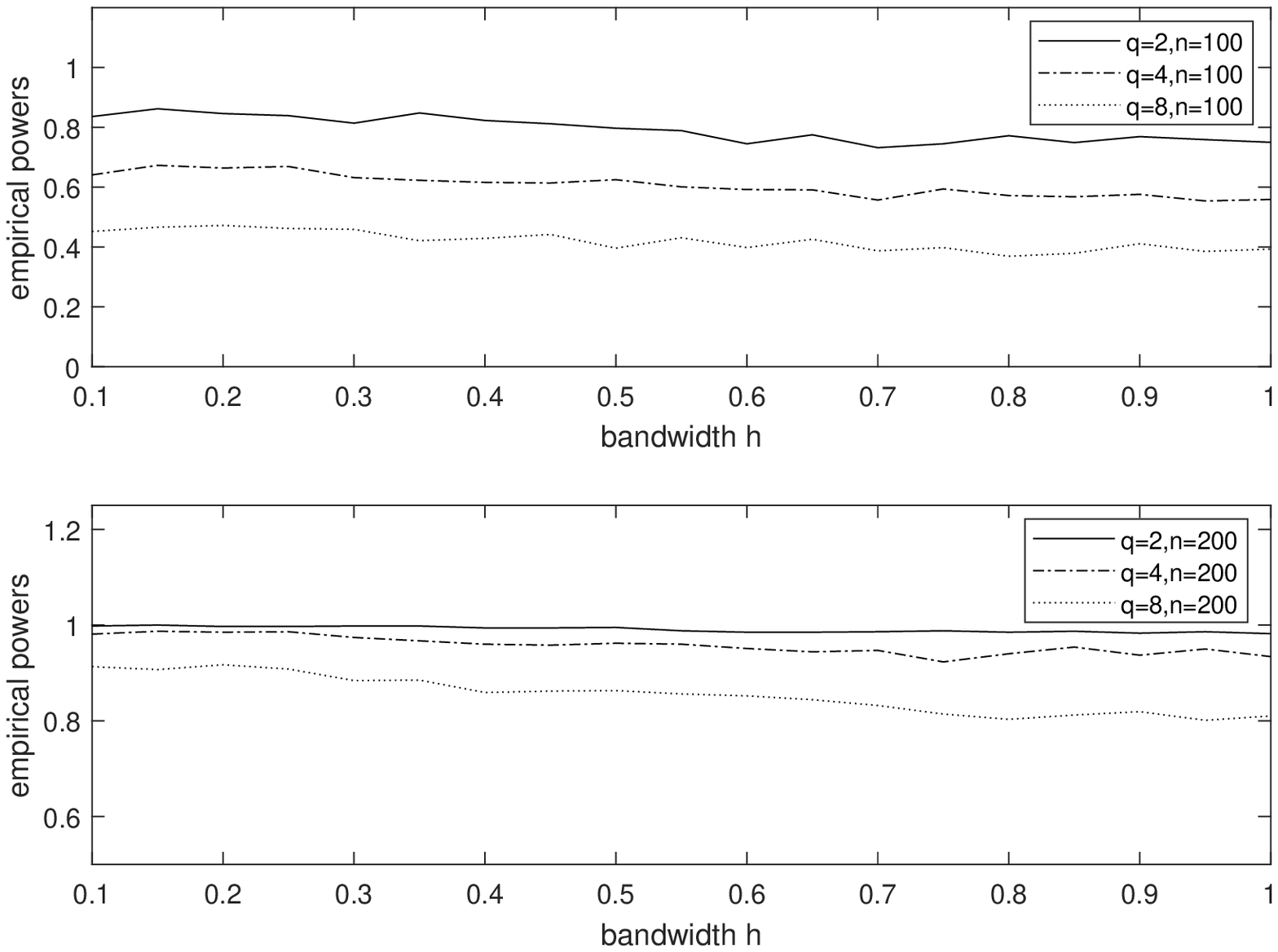}
  \caption{The empirical power curves of $HCM_n$ against the different bandwidths and sample size 100 and 200 with $a=0.2$ in Model~$H_{21}$.}\label{Figure 1}
\end{figure}

\begin{table}[ht!]\caption{Empirical sizes and powers of $HCM_n$, $T_n^{G}$, $T_n^{ZH}$, and $T_n^{ZFN}$ for $H_{21}$ and $H_{22}$ in Example 2.}
\centering
{\small\scriptsize\hspace{12.5cm}
\renewcommand{\arraystretch}{1}\tabcolsep 0.25cm
\begin{tabular}{cccccccccccc}
\hline
&\multicolumn{1}{c}{a} &\multicolumn{2}{c}{$HCM_n$} &\multicolumn{2}{c}{$T_n^{G}$} &\multicolumn{2}{c}{$T_n^{ZH}$} &\multicolumn{2}{c}{$T_n^{ZFN}$}  \\
&&\multicolumn{1}{c}{n=100}&\multicolumn{1}{c}{n=200} &\multicolumn{1}{c}{n=100}&\multicolumn{1}{c}{n=200} &\multicolumn{1}{c}{n=100}&\multicolumn{1}{c}{n=200} &\multicolumn{1}{c}{n=100}&\multicolumn{1}{c}{n=200}\\
\hline
$H_{21}, q=2$       &0.0           &0.044 &0.045   &0.047 &0.056   &0.044 &0.051   &0.035 &0.055\\
                    &0.1           &0.325 &0.671   &0.286 &0.513   &0.088 &0.210   &0.253 &0.747\\
                    &0.2           &0.771 &0.988   &0.685 &0.977   &0.283 &0.670   &0.636 &0.988\\
                    &0.3           &0.941 &1.000   &0.884 &0.998   &0.504 &0.892   &0.797 &0.994\\
                    &0.4           &0.970 &1.000   &0.956 &1.000   &0.664 &0.980   &0.840 &0.993\\
                    &0.5           &0.985 &1.000   &0.990 &1.000   &0.763 &0.995   &0.816 &0.990\\
\hline
$H_{21}, q=4$       &0.0           &0.040 &0.037   &0.059 &0.047   &0.027 &0.032   &0.025 &0.034\\
                    &0.1           &0.210 &0.527   &0.170 &0.290   &0.041 &0.093   &0.023 &0.274\\
                    &0.2           &0.583 &0.942   &0.434 &0.787   &0.100 &0.250   &0.086 &0.679\\
                    &0.3           &0.819 &0.991   &0.617 &0.965   &0.182 &0.425   &0.110 &0.732\\
                    &0.4           &0.895 &0.997   &0.750 &0.985   &0.254 &0.575   &0.144 &0.729\\
                    &0.5           &0.901 &1.000   &0.815 &0.996   &0.322 &0.681   &0.106 &0.664\\
\hline
$H_{21}, q=8$       &0.0           &0.042 &0.039   &0.065 &0.056   &0.046 &0.038   &0.024 &0.026\\
                    &0.1           &0.133 &0.330   &0.106 &0.160   &0.057 &0.048   &0.002 &0.003\\
                    &0.2           &0.409 &0.854   &0.190 &0.408   &0.066 &0.073   &0.000 &0.001\\
                    &0.3           &0.594 &0.966   &0.311 &0.660   &0.100 &0.116   &0.000 &0.003\\
                    &0.4           &0.736 &0.978   &0.388 &0.792   &0.130 &0.173   &0.001 &0.001\\
                    &0.5           &0.780 &0.979   &0.447 &0.858   &0.155 &0.200   &0.002 &0.006\\
\hline
$H_{22}, q=2$       &0.0           &0.046 &0.039   &0.052 &0.052   &0.024 &0.048   &0.039 &0.047\\
                    &0.1           &0.155 &0.327   &0.128 &0.207   &0.046 &0.087   &0.067 &0.235\\
                    &0.2           &0.477 &0.847   &0.375 &0.700   &0.138 &0.329   &0.248 &0.767\\
                    &0.3           &0.798 &0.995   &0.695 &0.973   &0.289 &0.693   &0.503 &0.955\\
                    &0.4           &0.937 &1.000   &0.893 &0.998   &0.521 &0.911   &0.640 &0.959\\
                    &0.5           &0.966 &0.999   &0.978 &1.000   &0.692 &0.978   &0.719 &0.941\\
\hline
$H_{22}, q=4$       &0.0           &0.040 &0.041   &0.059 &0.064   &0.041 &0.038   &0.033 &0.049\\
                    &0.1           &0.097 &0.220   &0.116 &0.142   &0.028 &0.056   &0.011 &0.051\\
                    &0.2           &0.328 &0.703   &0.230 &0.496   &0.065 &0.105   &0.023 &0.236\\
                    &0.3           &0.624 &0.977   &0.436 &0.809   &0.100 &0.232   &0.043 &0.437\\
                    &0.4           &0.807 &0.991   &0.686 &0.971   &0.203 &0.416   &0.055 &0.499\\
                    &0.5           &0.893 &0.988   &0.812 &0.997   &0.254 &0.593   &0.045 &0.430\\
\hline
$H_{22}, q=8$       &0.0           &0.044 &0.040   &0.068 &0.057   &0.051 &0.039   &0.033 &0.033\\
                    &0.1           &0.081 &0.120   &0.072 &0.078   &0.042 &0.053   &0.008 &0.002\\
                    &0.2           &0.229 &0.526   &0.132 &0.267   &0.081 &0.054   &0.001 &0.001\\
                    &0.3           &0.472 &0.877   &0.236 &0.478   &0.071 &0.104   &0.001 &0.001\\
                    &0.4           &0.662 &0.968   &0.329 &0.665   &0.115 &0.151   &0.000 &0.000\\
                    &0.5           &0.720 &0.944   &0.445 &0.843   &0.138 &0.206   &0.000 &0.000\\
\hline
\end{tabular}
}
\end{table}

\begin{table}[ht!]\caption{Empirical sizes and powers of $HCM_n$, $T_n^{G}$, $T_n^{ZH}$, and $T_n^{ZFN}$ for $H_{23}$ and $H_{24}$ in Example 2.}
\centering
{\small\scriptsize\hspace{12.5cm}
\renewcommand{\arraystretch}{1}\tabcolsep 0.25cm
\begin{tabular}{cccccccccccc}
\hline
&\multicolumn{1}{c}{a} &\multicolumn{2}{c}{$HCM_n$} &\multicolumn{2}{c}{$T_n^{G}$} &\multicolumn{2}{c}{$T_n^{ZH}$} &\multicolumn{2}{c}{$T_n^{ZFN}$}  \\
&&\multicolumn{1}{c}{n=100}&\multicolumn{1}{c}{n=200} &\multicolumn{1}{c}{n=100}&\multicolumn{1}{c}{n=200} &\multicolumn{1}{c}{n=100}&\multicolumn{1}{c}{n=200} &\multicolumn{1}{c}{n=100}&\multicolumn{1}{c}{n=200}\\
\hline
$H_{23}, q=2$      &0.0              &0.044 &0.049   &0.058 &0.057   &0.017 &0.034   &0.035 &0.067\\
                   &0.1              &0.101 &0.279   &0.143 &0.248   &0.061 &0.110   &0.126 &0.412\\
                   &0.2              &0.342 &0.739   &0.339 &0.688   &0.145 &0.350   &0.331 &0.853\\
                   &0.3              &0.606 &0.943   &0.615 &0.926   &0.276 &0.636   &0.642 &0.986\\
                   &0.4              &0.724 &0.992   &0.782 &0.992   &0.417 &0.855   &0.752 &0.999\\
                   &0.5              &0.844 &0.999   &0.867 &0.998   &0.555 &0.929   &0.831 &0.998\\
\hline
$H_{23}, q=4$      &0.0              &0.046 &0.041   &0.053 &0.048   &0.026 &0.024   &0.024 &0.015\\
                   &0.1              &0.096 &0.161   &0.106 &0.151   &0.039 &0.041   &0.015 &0.125\\
                   &0.2              &0.215 &0.528   &0.223 &0.442   &0.054 &0.110   &0.037 &0.439\\
                   &0.3              &0.368 &0.834   &0.342 &0.727   &0.089 &0.195   &0.118 &0.744\\
                   &0.4              &0.527 &0.935   &0.508 &0.873   &0.131 &0.316   &0.210 &0.846\\
                   &0.5              &0.652 &0.977   &0.598 &0.950   &0.187 &0.413   &0.233 &0.916\\
\hline
$H_{23}, q=8$      &0.0              &0.042 &0.048   &0.073 &0.050   &0.046 &0.046   &0.024 &0.028\\
                   &0.1              &0.059 &0.111   &0.085 &0.098   &0.048 &0.051   &0.008 &0.010\\
                   &0.2              &0.138 &0.331   &0.117 &0.192   &0.052 &0.048   &0.006 &0.007\\
                   &0.3              &0.244 &0.594   &0.185 &0.347   &0.066 &0.067   &0.001 &0.009\\
                   &0.4              &0.329 &0.741   &0.256 &0.479   &0.067 &0.072   &0.001 &0.009\\
                   &0.5              &0.403 &0.852   &0.270 &0.572   &0.081 &0.098   &0.003 &0.014\\
\hline
$H_{24}, q=2$      &0.0              &0.042 &0.038   &0.050 &0.051   &0.034 &0.038   &0.026 &0.045\\
                   &0.1              &0.035 &0.082   &0.074 &0.100   &0.052 &0.105   &0.055 &0.135\\
                   &0.2              &0.099 &0.211   &0.123 &0.258   &0.099 &0.286   &0.140 &0.546\\
                   &0.3              &0.169 &0.554   &0.230 &0.551   &0.182 &0.626   &0.327 &0.895\\
                   &0.4              &0.238 &0.719   &0.329 &0.769   &0.309 &0.813   &0.492 &0.966\\
                   &0.5              &0.277 &0.815   &0.392 &0.875   &0.343 &0.915   &0.580 &0.981\\
\hline
$H_{24}, q=4$      &0.0              &0.045 &0.037   &0.062 &0.064   &0.026 &0.032   &0.042 &0.044\\
                   &0.1              &0.041 &0.041   &0.069 &0.086   &0.035 &0.041   &0.022 &0.020\\
                   &0.2              &0.051 &0.093   &0.093 &0.153   &0.044 &0.059   &0.013 &0.073\\
                   &0.3              &0.077 &0.130   &0.136 &0.218   &0.047 &0.128   &0.020 &0.156\\
                   &0.4              &0.084 &0.225   &0.161 &0.352   &0.057 &0.159   &0.034 &0.242\\
                   &0.5              &0.095 &0.248   &0.180 &0.414   &0.084 &0.209   &0.040 &0.290\\
\hline
$H_{24}, q=8$      &0.0              &0.043 &0.041   &0.056 &0.052   &0.055 &0.040   &0.040 &0.037\\
                   &0.1              &0.046 &0.036   &0.077 &0.072   &0.047 &0.041   &0.021 &0.022\\
                   &0.2              &0.062 &0.051   &0.095 &0.099   &0.059 &0.045   &0.019 &0.007\\
                   &0.3              &0.054 &0.080   &0.110 &0.115   &0.043 &0.055   &0.008 &0.005\\
                   &0.4              &0.062 &0.115   &0.113 &0.147   &0.060 &0.076   &0.005 &0.008\\
                   &0.5              &0.080 &0.104   &0.123 &0.144   &0.071 &0.072   &0.003 &0.007\\
\hline
\end{tabular}
}
\end{table}

\begin{figure}
  \centering
  \includegraphics[width=12cm,height=4cm]{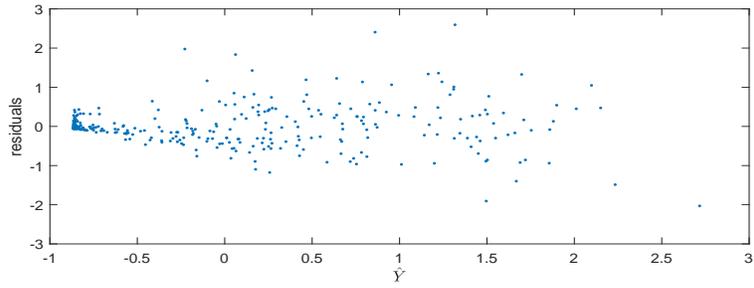}
  \caption{The scatter plot of the residuals $\hat{\varepsilon}_i$ against the fitted values $\hat{Y}_i$ for the baseball salary data set.}\label{Figure 3}
\end{figure}

\begin{figure}
  \centering
  \includegraphics[width=12cm,height=4cm]{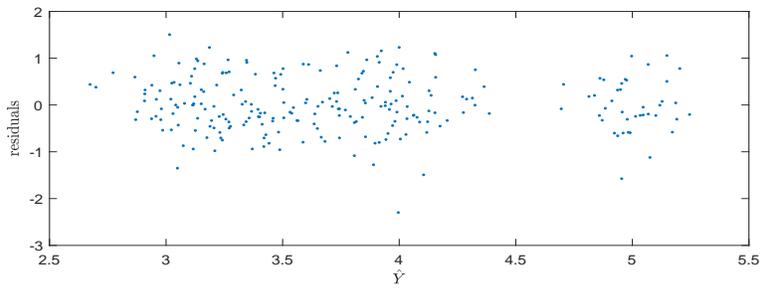}
  \caption{The scatter plot of the residuals $\hat{\varepsilon}_i$ against the fitted values $\hat{Y}_i$ for the ACTG 315 data set.}\label{Figure 4}
\end{figure}

\end{document}